\documentclass{article}

\usepackage{graphicx}

\newcommand{\jv}{\emph{JavaView}}

\graphicspath{{images/},{teaser/}}

\newtheorem{goal}{Goal}

\usepackage{hyperref}

\begin{document}

\title{How the deprecation of Java applets affected online visualization frameworks -- a case study}

\author{Martin Skrodzki, RIKEN iTHEMS\\ Wako, Saitama, Japan\\ mail@ms-math-computer.science}
	
	\maketitle
	
	\begin{figure}[h!]
		\centering
		\includegraphics[width=\linewidth]{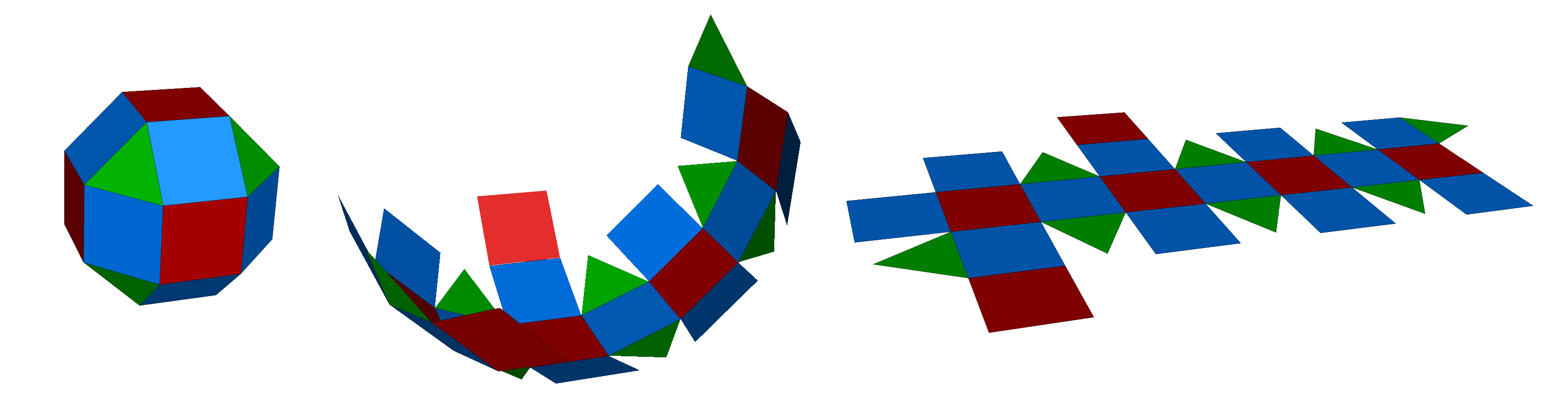}
		\caption{Unfolding an Archimedean solid with one of the mathematical web services of the \jv\ visualization framework.}
		\label{fig:teaser}
	\end{figure}
	\begin{abstract}
		 
		The \jv\ visualization framework was designed at the end of the 1990s as a software that provides---among other services---easy, interactive geometry visualizations on web pages.
		We discuss how this and other design goals were met and present several applications to highlight the contemporary use-cases of the framework.
		However, as \jv's easy web exports was based on \emph{Java Applets}, the deprecation of this technology disabled one main functionality of the software.
		The remainder of the article uses \jv\ as an example to highlight the effects of changes in the underlying programming language on a visualization toolkit.
		We discuss possible reactions of software to such challenges, where the \jv\ framework serves as an example to illustrate development decisions.
		These discussions are guided by the broader, underlying question as to how long it is sensible to maintain a software.  
	\end{abstract}


\section{Introduction}
\label{sec:Introduction}

At the end of the 1990s, the internet grew exponentially. More and more people discovered a multitude of options for using the new technology. In the context of science, it was investigated how education, publication of results, experiments, and general communication of scientific ideas could be realized online. A generally helpful tool for all of these applications is visualization. This holds particularly true for the realm of abstract mathematical content.

It is within this period that decisions where made to develop a new online visualization framework named \jv~\cite{JV}. It set out to tackle two major challenges that any mathematical visualization framework has to cope with. First, the mathematical content needs to be translated into a suitable visual setting. Second, the visualization has to be realized by the necessary technical steps in order to deliver it to the envisioned recipient. The situation at the end of the last century saw many researchers all over the world investigating visualizations of different mathematical objects and procedures. However, their works were generally published in the form of images or very short videos, where an interactive visualization would have been tremendously more informative~\cite[Sec.~1--3]{polthier2000mathematical}. A first goal of \jv\ was therefore to:
\begin{goal}[Interactivity]
	\label{goal:Interactivity}
	 Provide interactive, online visualization of mathematical content.
\end{goal}

Furthermore, as different research groups of the time tackled the implementation of visualizations and interactive applications, they all wrote their own code. These programs were tailored towards the specific hardware and operating systems available in the respective groups. Therefore, exchanging programs was not at all an easy task. Thus, a second goal of \jv\ was to:
\begin{goal}[Accessibility]
	\label{goal:Accessibility}
	Provide a system-independent framework that takes as much technicality away from the content creators as possible.	
\end{goal}
Thereby, the creators can focus solely on the production of new visualizations and research output. Furthermore, they can easily share the output based on the common framework. 

Finally, even the exchange of research data---like geometric models---was not easily possible because of the lack of a widely accepted and supported file format.  Therefore, regarding the field of experimentation in computational geometry and computer graphics, as a third goal, \jv\ should:
\begin{goal}[Communication]
	\label{goal:Communication}
	Introduce a unified file format for easy exchange of research and visualization data. 
\end{goal}
In fulfilling the goals outlined here, \jv\ competed with other contemporary frameworks, such as \emph{Cabri}~\cite{kuntz2002dynamic} or \emph{Cinderella}~\cite{kortenkamp2002interactive}. However, these two frameworks specifically tackle 2D visualizations, while \jv\ is to a large part concerned with interactive 3D applications. Still, the discussion on \jv\ in this paper is rather exemplary for these and other frameworks of the time. 

Guided by these three main goals, the authors of \jv\ aimed to create a framework that should be easily transferable between different architectures. Also, it should provide native export to web applications. To satisfy these constraints, the programming language \emph{Java} was chosen, with the included availability of \emph{Java Applets} for the internet, see Section~\ref{sec:ChoiceOfTheProgrammingLanguage}. In this language, a framework was developed for others to build on, see Section~\ref{sec:TheStructureOfJV}. After its release, extensive use was made of the new software and several widely used applications were created, some of which we discuss briefly in Section~\ref{sec:ApplicationExamples}.

However, after several successful years in the early 2000s, the \jv\ framework faced a significant challenge. The structure of \emph{Java Applets}---one of the core building blocks of \jv---was deprecated within the \emph{Java} language. Vendors of modern web browsers removed support for \emph{Java Applets} and banned them from their software, see Section~\ref{sec:JavaApplets}. This development left the \jv\ framework without one of its main features, i.e.\@ the easy web export. From this situation of the \jv\ framework as a basis, we start a broader discussion in Section~\ref{sec:ReactingToChanges} revolving around the question: What software frameworks can and what reasoning should this maintenance be based on? Results from the discussion and the entire article are summarized in the concluding Section~\ref{sec:Conclusion}.

\section{Choice of the Programming Language}
\label{sec:ChoiceOfTheProgrammingLanguage}

A first important decision in the creation of the \jv\ framework was choosing the underlying programming language. The contemporary popular high-level languages like \emph{Fortran}, \emph{C}, or \emph{C++} are all machine-dependent and require a compilation of the code on the specific machine of the user. Clearly, this hinders an easy distribution and dissemination of the framework. The \emph{Java} environment had been introduced in 1995~\cite{gosling1995java} and offered several appealing aspects towards the goals of the envisioned visualization framework~\cite[Sec.~4]{polthier2000mathematical}.

For instance, most of the contemporary browsers installed a \emph{Java} environment on the local machine. Hence, the potential user base---i.e.\@ those equipped with the necessary software---was almost equivalent to the users of the internet. Another particular advantage is that the installed \emph{Java} runtime environment (JRE) or virtual machine is independent of the actual browser software that uses it. Thus, all users had the same underlying runtime environment.

In terms of accessibility of interactive web content, \emph{Java} promised to be efficient in terms of data to be transferred. As the core classes and functionality of \emph{Java} are already present on the user's machine, these do not have to be downloaded. This makes the actual programs and applets extremely lightweight.

Furthermore, the build-in functionality of \emph{Java} comes with very accessible support of graphical user interfaces (GUI). As the programs are easily spread to other machines, operating systems, and users, a comprehensive GUI immediately became extremely important. That is, because a program with a series of convoluted text commands does not at all disseminate as well as a program with an intuitive GUI.

Finally, as stated above, the easy transfer of \emph{Java} programs to other machines and operating systems was a tremendous advantage of the new language. As the programs are not compiled into executable files, but bytecode, they can easily be executed on any target machine with a corresponding runtime environment or virtual machine. The developer therefore does not have to keep the specific architecture in mind any more when developing a program.

Additional to the programming language for the framework itself, also a new file format was sought for. This new format should enable easy exchange of created geometrical models and data. A basis for this format was found in the extensible markup language (XML), which had been introduced in 1998~\cite{bray1998extensible}. Based on this specification, the \emph{JVX} file format was developed. Being an XML-based format allows for easy automatic validation of \emph{JVX} files, which even works online~\cite{JVXValidator}. The format supports several geometric primitives as well as colors and textures~\cite{JVXFileFormatJV,JVXFileFormatMaple}.

These aspects of the new \emph{Java} language resonated well with Goals~\ref{goal:Interactivity} to~\ref{goal:Communication} set out for the \jv\ framework as described in Section~\ref{sec:Introduction}. Therefore, the decision was made to base the visualization framework on the \emph{Java} programming language.
	
\begin{figure}
	\includegraphics[width=1.\textwidth]{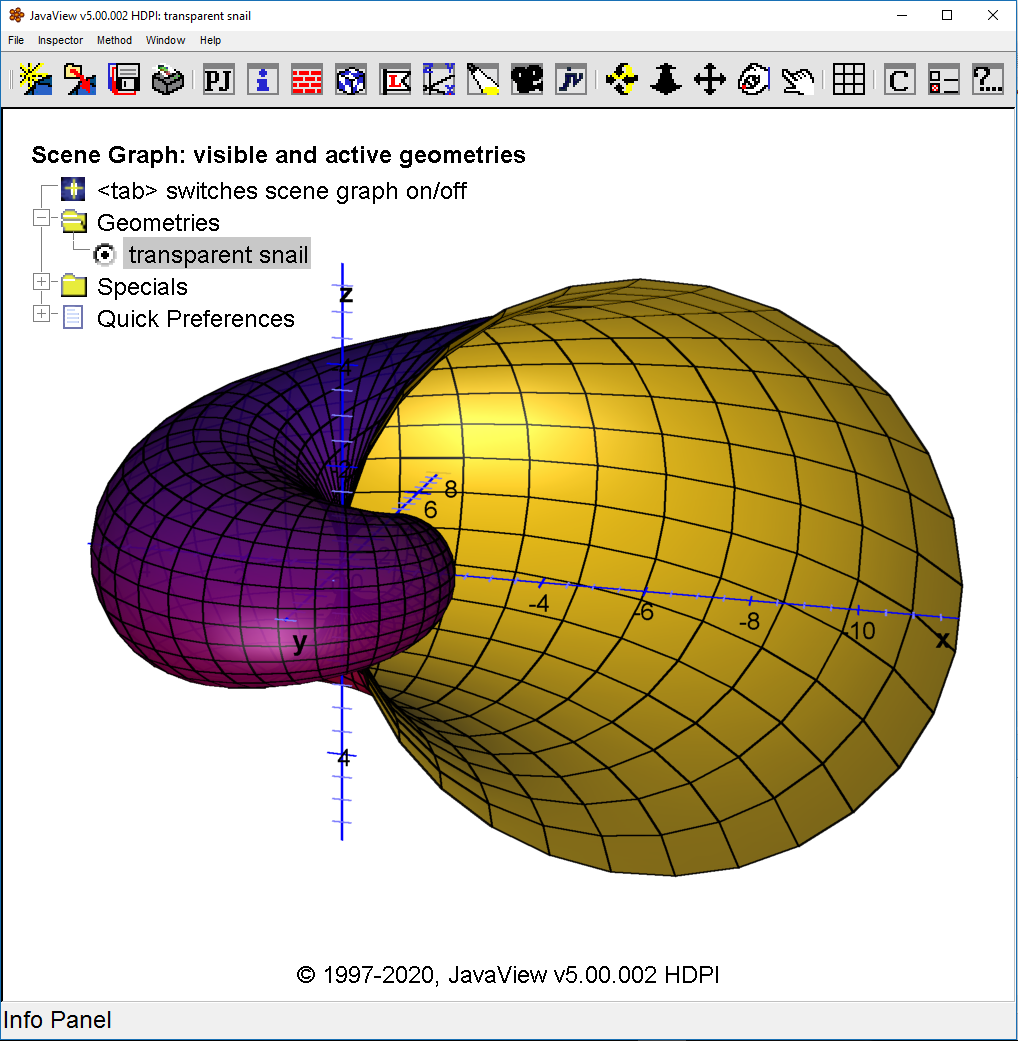}
	\caption{The \jv\ viewer with a logarithmic snail geometry.}
	\label{fig:jvViewer}
\end{figure}

\section{The Structure of \jv}
\label{sec:TheStructureOfJV}

After a one-year testing period at Technische Universit{\"a}t Berlin, the \jv\ software framework was first released in 1999~\cite[p.~1]{polthier2002publication}. Ever since, it was further developed at the Zuse Institute Berlin and at Freie Universit{\"a}t Berlin, where the current core development team is situated. The framework has always been a free software and can be downloaded on the corresponding web page~\cite{JV}. Since the release of version~3.0, a free yearly online registration is necessary to disable a ``missing license'' message in the viewer of the stand-alone version. Applets do not require a license. As of now, the code is not open-source, see Section~\ref{sec:ReactingToChanges} for a discussion of the implications of this decision.

The \jv\ framework comes with three main components~\cite[Sec.~2]{polthier2002publication}:
\begin{enumerate}
	\item \textbf{A software-based rendering engine integrated in a geometry viewer.} It supports basic functionality for exploration of geometries, comfort functions like coordinate axes and rulers as well as different camera modes. This functionality suffices for basic investigations of geometric models. 
	A general description of the capabilities of the viewer is available online~\cite{JVDocumentation}.
	\item \textbf{Different built-in \emph{workshops} and \emph{projects} for the creation and alteration of geometries.} Generally, \emph{workshops} apply to a shown geometry. The present functionalities allow for instance for mesh simplification, subdivision, smoothing, and optimization, among many other operations. Opposed to these \emph{workshops}, \emph{projects} implement more complex pipelines and can for example create their own geometries and animations. Among the long list of implemented \emph{projects} are Julia and Mandelbrot fractals, discrete minimal surfaces, polygonal curves, and different methods for handling and computing discrete vector fields on surfaces.
	\item \textbf{Class libraries and \jv\ archives for the development of new applications on the basis of \jv.} As the viewer and certain workshop functionalities are already present, the developer can focus on the creation of new applications and does not have to worry about technicalities. The libraries contain, for instance, data structures for geometry representations, algorithms for geometric modeling, numeric and linear algebra packages, animation frameworks, and support structures for user-interaction.
\end{enumerate}
Both the framework with its viewer and the corresponding \emph{JVX} file format were wide-spread in the contemporary Computer Algebra Systems. \emph{Maple} included \jv\ as a ``powertool'', while Mathematica, Matlab, and Polymake~\cite[Sec.~5]{polthier2002publication} as well as MuPAD~\cite{majewski2004using} supported or still support data exchange to and from \jv. Some of the listed programs also include the functionality to use \jv\ as a viewer of the generated data.

By providing a framework with the three components described above, \jv\ satisfied Goal~\ref{goal:Accessibility} outlined in Section~\ref{sec:Introduction}. Namely, developers and content creators do not have to deal with technicalities like a rendering engine or a geometry viewer. Also, they do not have to re-write basic geometry processing algorithms, as those are readily available. Thus, they can start immediately with programming and creating their own mathematical visualization projects. Furthermore, the spread of the \emph{JVX} file format contributed to an easy exchange of geometric data between different systems, which works towards Goal~\ref{goal:Communication}.

Finally, another large benefit of \jv\ was the easy creation of interactive web visualization via \emph{Java Applets}~\cite[Sec.~5.2/5.3]{polthier2000mathematical}. In the following, we will elaborate on this functionality by discussing some examples that highlight the broad applicability of \jv\ as well as the different areas in which it was utilized.

\section{Application Examples}
\label{sec:ApplicationExamples}

The applications that are presented here are described in the way they were published, i.e.\@ at the technological stage of their respective release date. We discuss the problems that these different software modules tackled. Keep in mind the current, advanced status of the \jv\ framework, as referred to in Section~\ref{sec:ReactingToChanges}.

\subsection{Knot Simplifier}
\label{sec:KnotSimplifier}

\begin{figure}
	\includegraphics[width=1.\textwidth]{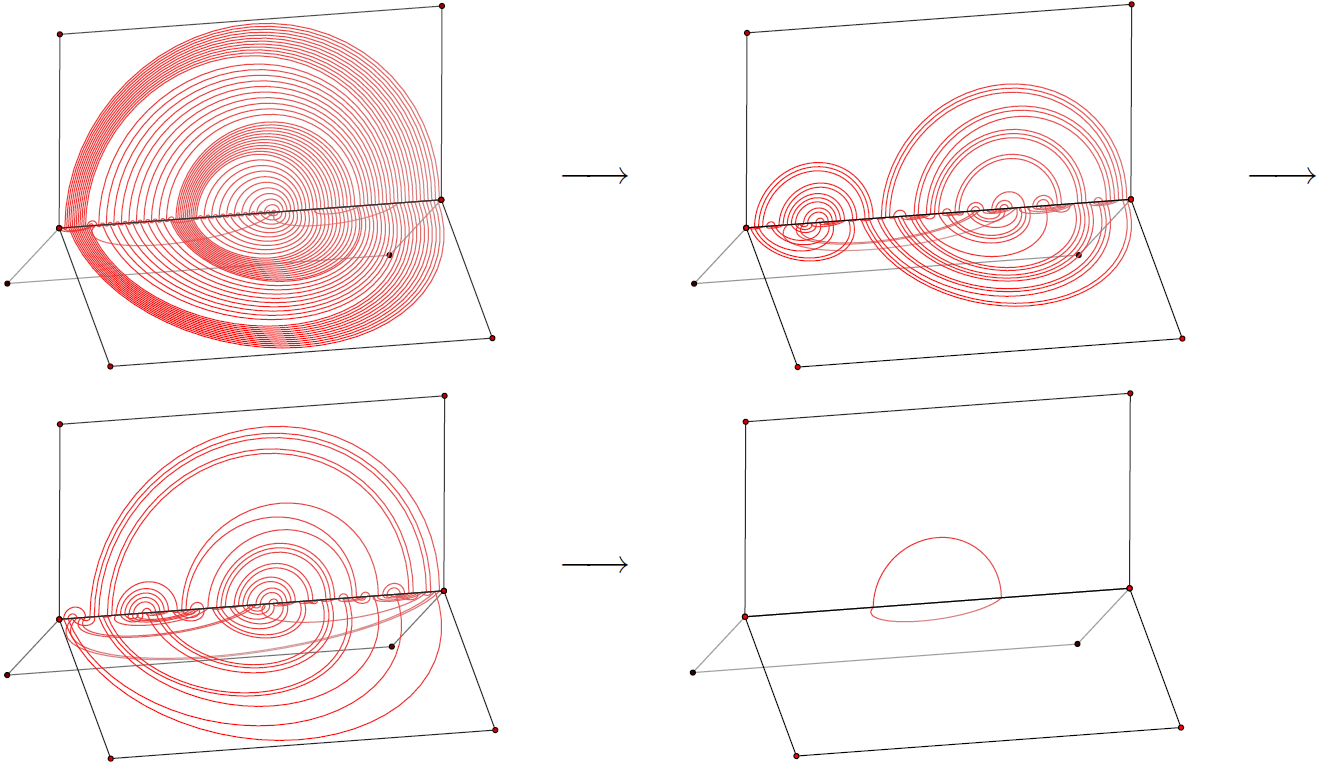}
	\caption{Procedure of ``untangling Goeritz 13-21'', visualized with \jv, taken from~\cite{andreeva2002mathematical}.}
	\label{fig:unknot}
\end{figure}

The authors of~\cite{andreeva2002mathematical} present the implementation of a partial knot recognition algorithm. It stands aside from other algorithms in the field as it is implemented in the form of a web service on the basis of \jv, available at~\cite{JVServices}. The applet allows users to view knots from an online database, create new knots, and (partially) simplify a knot. See Figure~\ref{fig:unknot} for a visualization of the procedure.

This interactive applet for the discovery of knots is a representative of several mathematical online services implemented via \jv~\cite{JVServices}. Aside from the knot simplifier and a simple geometry viewer, the list includes an ODE solver, a root finder for functions, a visualization of geodesics on polyhedral surfaces, an algebraic solver, and a mesh unfolder that provides a planar net for a given geometry, see Figure~\ref{fig:teaser}. All these applications are available in form of \emph{Java Applets} online. Thus, satisfying Goal~\ref{goal:Interactivity} from Section~\ref{sec:Introduction}, these services provide a low-threshold means to interact with and learn mathematics online.

\subsection{\emph{Maple} and \emph{JavaViewLib}}
\label{sec:MapleAndJavaViewLib}

\begin{figure}
	\includegraphics[width=1.\columnwidth]{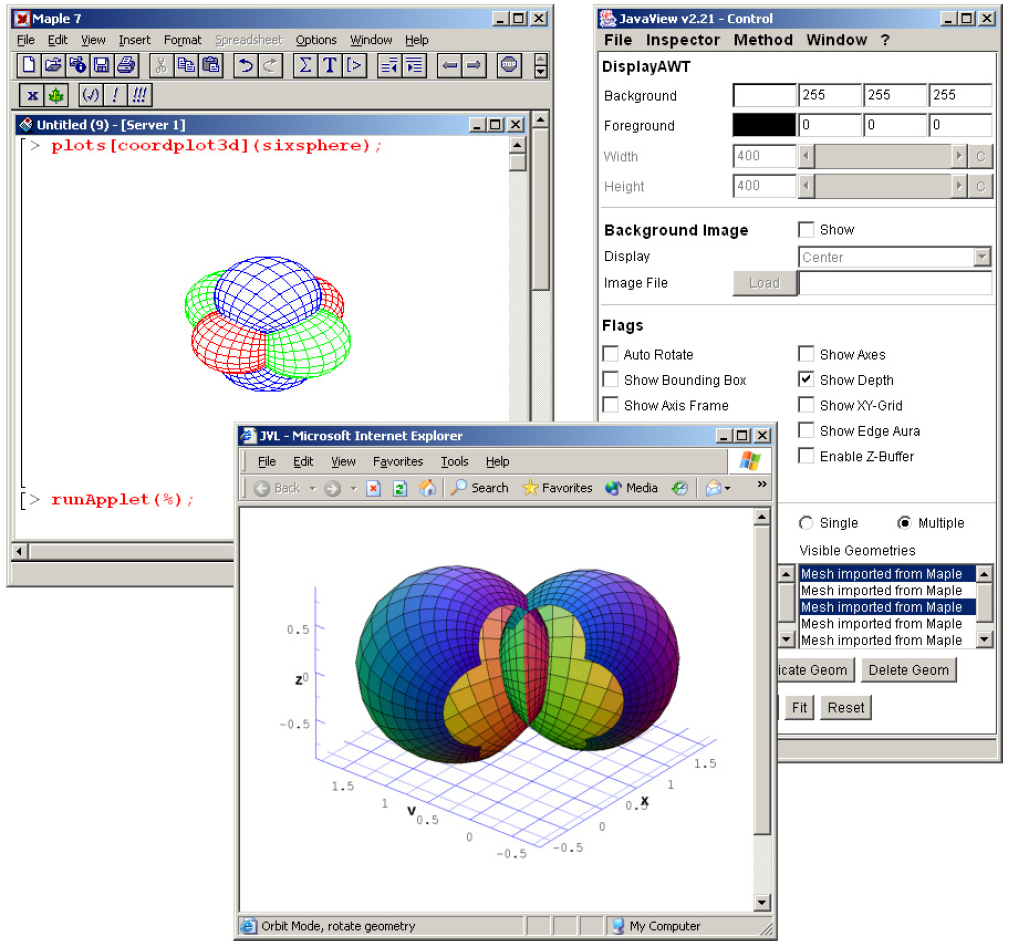}
	\caption{The \emph{JavaViewLib} enables the creation of an interactive web page out of any \emph{Maple} plot, taken from~\cite{dugaro2003visualizing}.}
	\label{fig:maple}
\end{figure}

The \emph{JavaViewLib} enables full support of \jv\ from within \emph{Maple} in the form of a ``power tool''. This tool adds new interactivity to \emph{Maple} plots in both web pages and worksheets. For instance, it introduces the widely used arc-ball rotation system~\cite{shoemake1992arcball} to \emph{Maple}. Arguably the most powerful addition that \jv\ provides for \emph{Maple} comes in the form of an easy web page export functionality. By a simple command, a complete \emph{Maple} plot can be exported into an \emph{HTML} document with \emph{Java Applets}, allowing for interactive exploration of the plot~\cite{dugaro2003visualizing}, see Figure~\ref{fig:maple}.

In this application, \jv\ enables the users of \emph{Maple} to easily share their work and make it more accessible, in particular via the internet. In terms of Goal~\ref{goal:Communication} formulated in Section~\ref{sec:Introduction}, this serves the general communication of scientific ideas as researchers can easily share, alter, and work on different geometric models.

\subsection{EG-Models}
\label{sec:EGModels}

\begin{figure}
	\centering
	\includegraphics[width=0.95\textwidth]{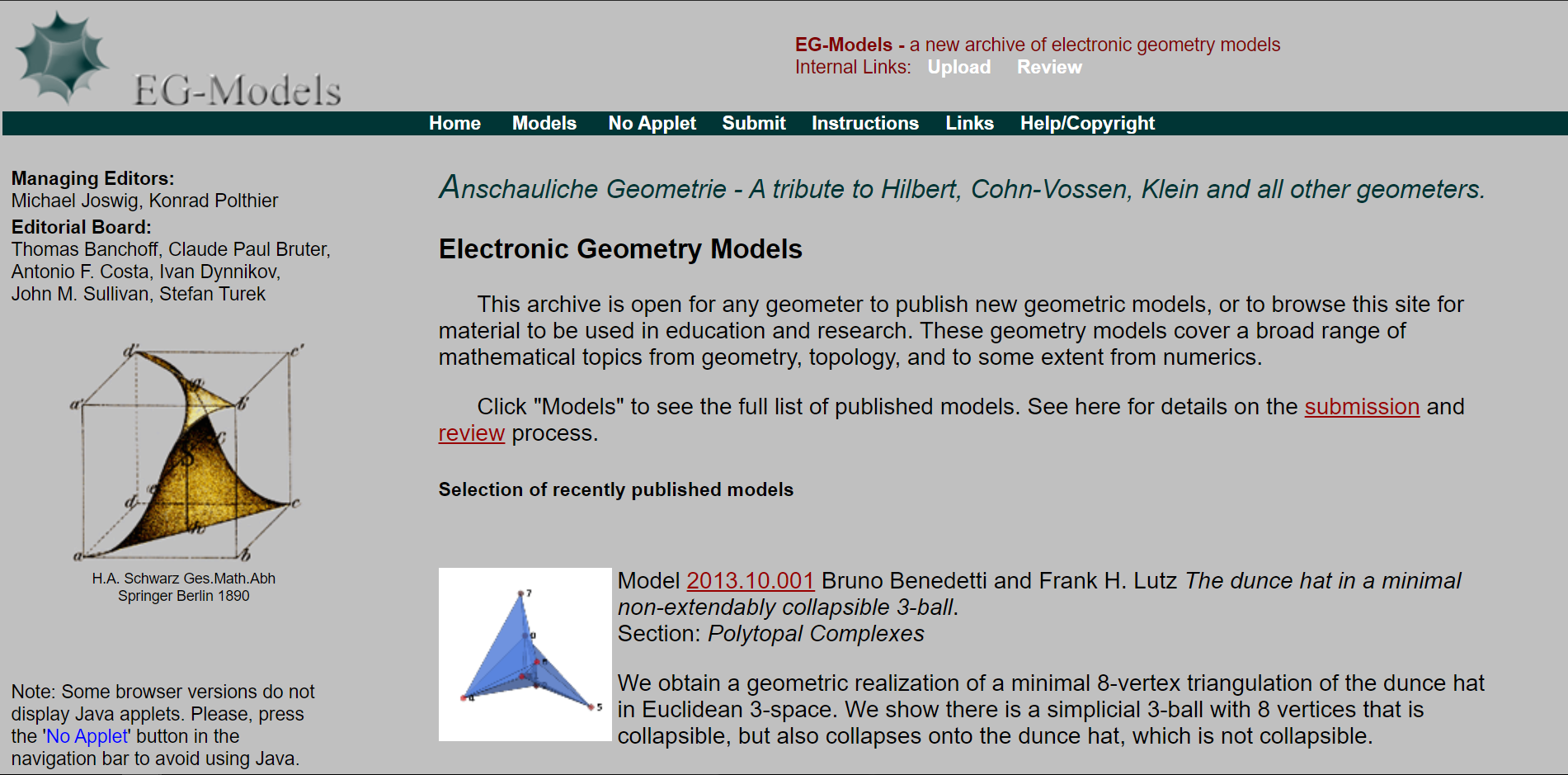}
	\caption{Screenshot of the EG-Models web page~\cite{joswig2002eg}.}
	\label{fig:eg-models}
\end{figure}

Another example for the use of \jv\ is the online journal for electronic geometry models, short ``EG-Models''~\cite{joswig2002eg}. It aims to exhibit a broad collection of peer-reviewed geometry data sets from a wide range of mathematical subjects such as---but not limited to---differential, discrete, or computational geometry. Aside from images and interactive visualizations via \jv, the key aspect is the data itself, combined with a self-contained description of its mathematical importance. See Figure~\ref{fig:eg-models} for a screenshot of the web page.

The ``EG-Models'' project creates a whole new outlet for scientific research. In the form of a peer-reviewed online journal, it offers a possibility for publication of geometrical data. At the same time, the website provides an accessible way of browsing through the data interactively within a \jv\ applet. This enables researchers to use these curated model sets in their own research by simply browsing them interactively on the web page and by downloading those that are helpful for their own projects. In terms of Goal~\ref{goal:Communication} from Section~\ref{sec:Introduction}, the ``EG-Models'' web page thus offers a new way to publish research work.

\subsection{Geometry Processing}
\label{sec:GeometryProcessing}

The description of the workshop and projects in Section~\ref{sec:TheStructureOfJV} already hinted at the capabilities of the \jv\ framework regarding general geometry processing tasks. While the previous applications where concerned with different interactive web pages, the most powerful algorithms are available in the \jv\ stand-alone program. Most notably, this offline version of \jv\ can handle all relevant aspects of the geometry processing pipeline---starting from  scanned, real-world models, processing them, and preparing them for 3D printing. For instance, isotropic and anisotropic smoothing algorithms are available to remove noise components added during the scanning process. Boundaries can be identified and corresponding wholes can be filled automatically in order to create a watertight surface. A mesh can either be subdivided via different schemes or simplified, depending on the current application. Furthermore, it can be altered according to the minimization of different energies, like Dirichlet or conformal stresses. Finally, several extrusions are possible to prepare the object for 3D printing. The framework supports output to eleven different 3D geometry formats, including the widely used STL format for 3D printing. 

\subsection{Further Applications}
\label{sec:FurtherApplications}

Aside from the applications presented above, further examples for the integration and application of \jv\ can be found in print~\cite[Sec.~4]{polthier2002publication} as well as online~\cite{JVServices,JVApplications}. Finally, \jv\ has been discussed in the specific context of use in schools~\cite[Sec.~4]{majewski2004using}.
	
-------------------------------------------------------------------------
\section{Changes in the Programming Language -- Deprecation of Java Applets}
\label{sec:JavaApplets}

As discussed above, one main feature of the \jv\ framework is its easy export of interactive mathematical visualizations to web pages. This functionality depends on the technology of \emph{Java Applets}. Thus, the development of this aspect of the \jv\ framework is dependent on the development of \emph{Java Applets}.

An online article by Michael Byrne provides a well-written summary of the history of \emph{Java Applets}~\cite{byrne2016rise}. While the technology of applets was tremendously successful in the early 2000s, the come-back of \emph{JavaScript} (JS) brought a serious competitor back into the field. The \emph{Chrome} browser soon supported JS with its own engine, making the additional installation and frequent updates of \emph{Java}'s virtual machine a comparable hassle for the user. Several exploits and security faults gave \emph{Java Applets} a bad standing in the community, as Ben Evans summarized in his 2015 book~\cite{evans2015java}.
Shortly after the book was issued, the release of \emph{Java} version~9.0 saw the deprecation  of the \emph{Java Applet} API, ``as web-browser vendors remove support for Java browser plug-ins''~\cite{titov2016jep}. After this preliminary step, the support for the \emph{Java Applet} API was completely removed in the release of the \emph{Java} Development Kit (JDK), version~11.0~\cite{JDK11}.
	
With the end of support for \emph{Java Applets} by all major web browsers, the easy web export from the \jv\ framework was gone. In particular, this caused the web services~\cite{JVServices} as well as the applications as discussed in Sections~\ref{sec:KnotSimplifier}--\ref{sec:EGModels} to become unavailable with modern browsers online. They are now only accessible via the stand-alone offline version of \jv. Therefore, a serious reorientation of the framework, its main applications and goals was necessary. Certainly, given the amount of time, energy, and resources that went into the algorithms, implementation, and general framework concepts, a reorientation should build on this basis and preserve this work, if possible. We will discuss different aspects of this reorientation in Section~\ref{sec:ReactingToChanges}. Here, we briefly consider how different software handled the historical development. 

The contemporary frameworks \emph{Cabri} and \emph{Cinderella} had to face similar challenges as \jv. When considering how the \jv\ framework might react and change, it can be beneficial to consider the actions taken within these two frameworks. They followed different approaches. The \emph{Cabri} software nowadays focuses on education. It runs in a stand-alone environment on PC or Mac and targets both teachers and students~\cite{Cabri}. Thus, it moved away from its research background and applicability in web pages. Instead, it became a user-centered, commercial software.

The original stand-alone version of \emph{Cinderella} is still up and running. It is actively worked on and recently became freely available after 20 years of commercial distribution~\cite{Cinderella}. However, the interactive examples on the website are only accessible with a web browser supporting \emph{Java}, i.e.\@ not with any modern web browsers. To counteract the loss of the created material, a sister-project to \emph{Cinderella} was formed, called \emph{CindyJS}~\cite{gagern2016cindyjs}. Recognizable from the name, it is implemented in \emph{HTML5}, JS, and \emph{WebGL}. Therefore, it is suited to be a replacement of the \emph{Java Applets} used by \emph{Cinderella}. As \emph{CindyJS} provides compatibility to existing \emph{Cinderella} projects, these can now easily be recycled and be given a fresh start as lightweight and interactive web applications. See for instance the examples in the \emph{CindyJS} gallery~\cite{CindyJSGallery}.

The \jv\ framework is facing the challenge that parts of its underlying technology are deprecated and it cannot provide one of its main features anymore. Considering other contemporary frameworks, it can be shown that not only changes in the software environment, but a variety of different factors can be a challenges to a specific software. \emph{Cinderella} and Cabri reacted to their respective challenges in their own ways. The following section is devoted to general thoughts about such reactions, always considering \jv\ as the element of this case study.

\section{Developing a Software Framework over Time}
\label{sec:ReactingToChanges}

In the sections above, we have seen how the \jv\ framework was impacted by the deprecation of \emph{Java Applets} that formed the basis for one of its main functionalities, the easy web export. For the remainder of the article, we will broaden the view and consider a more general question in the field. Namely, phrasing it trenchantly, whether or not all software should be saved. We will approach this question via various facets and always refer back to the case study of \jv\ to give concrete examples for respective reactions. The different facets of the question are not necessarily disjoint, but rather overlap and affect each other.

First, however, we need to define what ``saving'' a software means in the context of the following discussion. Certainly, there is a broad range of states a software can be in. It can be under active development by one or several developers; it can be executable on modern hardware after installation, while not being further maintained; it can be available in an archived or legacy format, for instance within a container, equipped with other, necessary pieces of software not commonly available anymore; or it can be non-executable in any form on modern hardware. While the latter definitely describes a piece of `dead' software, the other cases are more subtle. We will assume in the following that a software is safe in securing its own existence only if its actively maintained by developers. Otherwise, one could argue, it is doomed to fall into the last category of eventually becoming non-executable. Following this definition, the \jv\ framework is actively maintained and thus ``saved'' by its developer team. Despite its long lifetime since its first release in 1999, it saw the release of version 5.01 in March 2020.

We have to take care, however, not to take a long lifetime of a software as a valid reason for saving it. While it might hurt to toss aside a large collection of experience, research, algorithms, and implemented concepts, not all software can be saved. Otherwise, a field would become too fragmented to actually accomplish anything. Therefore, it is important and necessary to frequently check the applications and thus the raison d'{\^e}tre of each software framework.

Before we dive into the discussion of the different facets that contribute to the development of a software framework over time, we will briefly form the ground of this discussion by collecting reasons and circumstances that require a software framework to change in the first place. An important factor are ever-changing restrictions and limitations. Without claiming completeness, the following factors limit the development of a software framework.
\begin{itemize}
	\item A given setup can only achieve a level of \textbf{security} according to the weakest link in its chain. A major reason for the deprecation of \emph{Java Applest} was their security issues, in combination with a growing desire by the users to have secure web application.
	\item The \textbf{programming language} of the software provides certain functionalities, but also comes with constraints. For instance, \emph{Java} is easily portable between different architectures, but has a bad performance when compared to e.g.\ \emph{C++}.
	\item \textbf{Portability} is a factor that is also affected by the rise of new architectures. While the scientific computing sector is currently turning towards \emph{Python}, the \emph{Java} language has witnessed a readjustment with the \emph{Android} system for mobile devices.
	\item \textbf{Performance} is mostly a question of the desired use case. In terms of Goal~\ref{goal:Interactivity}, \jv\ should load fast and be extremely interactive, without lags or long waiting times. These aspects are less important, when focusing on other applications, like scientific geometry processing, see Section~\ref{sec:GeometryProcessing}.
	\item Similarly, the ease of use of a software comes with its \textbf{user base}. Again, educational and interactive visualizations need to have more accessible and intuitive graphical user interfaces than highly professionalized software for rather specific tasks.
\end{itemize} 
Evolutionary speaking, these restrictions provide niches into which software with their respective aims and use cases can nest. Another important factor are changes in all aspects of a software framework that possibly affect the limitations as well as the decisions made in these regards. For instance, over time, support for old hardware is given up in favor of new developments. Similarly, software and their underlying languages also evolve and wherever old aspects vanish, new possibilities and opportunities arise. In the following, we will discuss several such changes and possible reactions.


\subsection{How to react when a main use case of an application is put to the test?}

When a software loses one of its main use cases---like the easy web export functionality of \jv---a thorough reorientation is necessary to ensure that new goals and target areas are identified and focused on. At the point of deprecation of \emph{Java Applets}, the \jv\ framework was a mixture of a viewer, some core functionality, and several high-level applications/algorithms implemented on the basis of the other two. A development decision based on this status breaks down into the question: What should the main goal of the software framework in the future be?

In case the creation of web applications is the main goal of the framework, corresponding measures would have to be taken to find a replacement for the deprecated \emph{Java Applet} technology. Furthermore, these web applications could either provide mathematical illustrations or be more complex and therefore rather mimic the mathematical web services~\cite{JVServices}. In a way, the \emph{CindyJS} project~\cite{gagern2016cindyjs} followed the first path and now provides a set of mathematical illustrations via its gallery~\cite{CindyJSGallery}. Regarding the second aspect of more complex web applications, the \emph{Visualization Toolkit} (VTK)~\cite{schroeder2004visualization} expanded its functionality via a JS add-on, called VTK-JS~\cite{VTKJS}.

Regarding the \jv\ framework and its components as discussed in Section~\ref{sec:TheStructureOfJV}, when \emph{Java Applets} were deprecated, the development team decided to continue work on the stand-alone version of the software and its included algorithms. Thereby, the \emph{MeshLab} software became an immediate competitor~\cite{cignoni2008meshlab}. However, \jv\ still has its class libraries available to those users who would like to expand and alter the software towards their own use cases. This gives \jv\ a slight benefit over \emph{Meshlab}. Conversely, \emph{Meshlab} is implemented in \emph{C++} and thus provides better performance compared to \jv\ while still being available for all major platforms.

This shows that when a software is confronted with loosing one of its main use cases, it has two options. It can either make all necessary changes and additions to still follow this application, like \emph{CindyJS}, or it can drop this use case and focus on remaining application scenarios, like \jv.


\subsection{What aspects of software can be maintained by container technology?}

In the preliminary definition of ``saving'' software, we have deemed container technology as an unsuitable way for saving a framework. However, this broad view cannot be upheld when considering certain, more specialized use cases. Consider for instance a set of visualization experiments that are programmed and executed as supplementary data for a research article. Naturally, subsequent articles will cite these experiments and will strive to compare them to their own results. In this sense, it is important to provide an executable environment for the computations, even if the underlying software setup changes. The field of container technology aims at solving exactly this problem. By using available techniques such as \emph{Docker} or \emph{VirtualBox}, it is possible to provide an out-of-the-box running environment even for outdated software. Note that this general approach of keeping experiments available and thus replicable is the main goal of the \emph{Graphics Replicability Stamp Initiative}~\cite{ReplicabilityStamp}.

In terms of \jv, container technology will not be able to help maintaining the framework. This still needs to be done by active developer teams. However, a container can include a JDK, a browser, and a \emph{Java Applet} (e.g.\@ from the online web services~\cite{JVServices}). If the JDK and the browser are provided in versions that still support \emph{Java Applet} technology, container can be a means to still provide the discussed services with a comparably low effort. Such container could be made available in the relevant repositories as well as via download from the official \jv\ website~\cite{JV}. Thereby, research process is still accessible and follow up works can benefit from it.


\subsection{How does the choice of the programming language affect a software framework?}

The discussion in Section~\ref{sec:ChoiceOfTheProgrammingLanguage} has shown that choosing a programming language for a software framework is based on the respective goals to follow. However, what happens if the language changes and does not provide the elements for these goals anymore? Certainly, the underlying programming language evolves and corresponding problems and the necessity for adjustment will occur. Is the translation into another language a valid option to handle these challenges or can other features of the original language be employed to circumvent the problems?

Regarding the \jv\ software framework, an obvious choice was to leave the implementation in \emph{Java}. While this circumvents the refactoring or complete rewriting of the code, it also---given the deprecation of \emph{Java Applets}---implies that all functionality of the framework based on \emph{Java Applets} becomes inaccessible. Nonetheless, it is a reasonable choice, as the once popular \emph{Java3D} library is no longer officially supported, which provides \jv\ with a unique characteristic in the \emph{Java} domain. Within this field, \jv\ now is the only software framework to provide all relevant aspects of a visualization toolkit, including viewer, core functionality, and algorithms. This still comes with the benefits of the \emph{Java} language, which has a low threshold in terms of its setup such that---in particular user with few experience---can tackle programming projects faster than in e.g.\ \emph{C++}.

Other languages---aside from \emph{Java}---do have widely used geometry frameworks available. Popular examples include the \emph{CGAL}~\cite{cgal2020cgal} framework or the open-source software \emph{Meshlab}~\cite{cignoni2008meshlab}, both implemented in \emph{C++}. These large collections provide support for viewing operations as well as core functionality. Therefore, they are ideal for focusing on the implementation of high-level algorithms or application and services, just like \jv\ does in the \emph{Java} domain. Similarly, several 3D geometry packages exist for more specific purposes (like the point cloud library for processing of unstructured point sets~\cite{rusu20113D}) or within other languages (in \emph{Python} for instance the bindings for tetrahedral meshing~\cite{hu2018tetrahedral} or the packages for geometric algebra~\cite{clifford}).

Coming back to the original roots of \jv---interactive web page applications---several other projects have now filled the gap. The aforementioned \emph{CindyJS}~\cite{gagern2016cindyjs} as well as VTK~JS~\cite{VTKJS} both rely on a larger framework in the background and only provide a JS interface to make this background framework accessible in the web. Other approaches, like \emph{three.js} enable the user to render 3D web applications using \emph{WebGL}, thus providing less comfort, but more flexibility. These examples show that combinations of different languages are possible within the domain of a single framework and that \jv\ might at some point consider using an export into web applications, not based on \emph{Java Applets}, but possibly on one of these existing approaches.


\subsection{What other basic building blocks of a software framework are subject to change?}

As indicated in the beginning of this section, several different factors and limitations contribute to the development of a software. In particular a visualization framework has to cope with more than changes in the programming language and related software components---like the operating system. Other changes equally affect the performance of the framework. Consider the following three examples in the realm of the \jv\ framework.
\begin{enumerate}
	\item The wider availability of potent hardware in the user base creates the expectancy of the users that the visualization framework of their choice also supports this hardware. Towards this end, \jv\ has---aside from its software rendering---added support for \emph{OpenGL}, thereby harnessing the graphics power at the user's machines, but still remaining platform independent as \emph{OpenGL} is available for a variety of platforms.
	\item With rising graphic capabilities, the resolution of the user's monitors are also on an incline. To pick up this movement, recent versions of \jv\ come with a high DPI mode to better scale on systems with 4k resolution.
	\item Not only output devices, but also input devices develop. Aside from traditional input via a mouse with the arcball model~\cite{shoemake1992arcball}, new input devices like the \emph{Leap Motion Controller} allow the user to interact with the programs using hand gestures. A corresponding support has been added to the \jv\ framework recently~\cite{skrodzki2019leap}.
\end{enumerate}
Missing reaction to these changes easily results in frustration on the side of the users who then migrate to other systems that better satisfy their demands. Therefore, each framework offers its respective attractive benefits. For instance, \emph{CGAL}~\cite{cgal2020cgal} provides native multi-core support, while \emph{CindyJS}~\cite{gagern2016cindyjs} offers easy, encapsulated access to harness the computation power of the graphics card.


\subsection{How does a closed- or open-source policy affect the development of a framework?}

The availability of software as open-source clearly affects the number of available developers, collaborators, and users willing to interact with the software. However, a closed-source software with a loose release cycle allows for closer monitoring of the developed functionality and can provide better quality and more homogeneous code. Also, all project members can focus on the continued development and no resources have to be allocated towards code reviews of external collaborators. As the \jv\ framework is currently closed-source, developments and additions made by the users outside of the core development team are not integrated into the software.

Other software packages in the field tend to have a policy between the two extremes of open- or closed-source. The \emph{CGAL} project~\cite{cgal2020cgal} for instance has an editorial board. Additions made to the software have to pass the board before they are included in the main framework. The driving force for \emph{CGAL} are its industry clients who propose and require new features to be added. This provides a natural selection for the features that are contributed and considered for inclusion in the new releases.

Both \emph{Meshlab}~\cite{cignoni2008meshlab} and \emph{CindyJS}~\cite{gagern2016cindyjs} handle contributions via pull requests to their respective repositories. However, when considering the contribution statistics for these frameworks, they are still largely carried by a small number of core-developers with the majority of the users providing minor contributions. It remains debatable and dependent on the actual framework whether a code review of contributed code from outside the core developer team takes more or less time compared to the core developers creating a solution for the same issue.


\subsection{How to adjust to competitors coming into the field?}

As new developments arise in a field, new niches open and new competitors rush in to fill these. Established visualization frameworks thus have to cope with other software coming in and competing for the same user base. In the case of \jv, several other developments have picked up possibilities for easy web export after the deprecation of \emph{Java Applets}. For instance, \emph{three.js} is a framework to render 3D web applications using \emph{WebGL}. Several visualizations are available based on this technology. Furthermore, the \emph{Unity} framework makes it very simple to create interactive setups that can be exported to a variety of targets, like web pages, game consoles, or mobile applications. The commercial, scientific computing software \emph{Matlab} has a web export via its \emph{Simulink WebView} functionality, \emph{Cinderella} projects are made available online via the discussed \emph{CindyJS}, and \emph{Python} visualizations can be shared online in the form of \emph{Jupyter Notebooks}. Even more recently, the \emph{Java}-style language \emph{Processing} causes a wide spread of online visualizations and interactive displays.

As these competitors have rushed into the field of online visualizations, \jv\ has concentrated on its stand-alone program and the corresponding implemented high-level geometry processing algorithms (see Section~\ref{sec:GeometryProcessing}). In this area, it competes with current libraries, like \emph{CGAL}~\cite{cgal2020cgal} or \emph{libigl}~\cite{libigl}. Other libraries focus on algorithms for specific application areas, like the \emph{Point Cloud Library} which aims to process unstructured point sets~\cite{rusu20113D}. While these three libraries are working with \emph{C++}, comparable options are available in other languages, such as \emph{Python}, cf.\@ the aforementioned~\cite{hu2018tetrahedral,clifford}. While libraries are mainly aimed at larger setups into which they can be integrated, the \emph{C++}-based \emph{Meshlab} program provides a stand-alone setup with a user-friendly graphical interface to process geometric models~\cite{cignoni2008meshlab}. Less language-dependent, the \emph{ParaView} software offers developer support in \emph{C++}, \emph{Python}, and a web version via JS~\cite{ahrens2005paraview}.

Given the developments in the field of visualization software, \jv\ has to continuously readjust and check its raison d'{\^e}tre against the competitors. Currently, it is the only available and actively maintained visualization framework in the \emph{Java} language, which provides its unique attraction. It has successfully managed to adapt to different situations over the course of the last 21 years and thus sets an example by its development choices for other frameworks facing similar challenges.


\section{Conclusion}
\label{sec:Conclusion}

In this article, we have presented the history of the visualization framework \jv.
Originally, the software tackled the specific problem of interactive geometry visualizations in the internet as one of its main applications.
This feature was implemented on the basis of \emph{Java Applets}, which became deprecated in 2016.
Having lost the availability of web exports, the \jv\ framework had to reorganize and readjust itself within a market of different goals, competitors, and user demands.

As the \jv\ software has existed and been maintained for 21 years, it has several valuable lessons to tell. 
Much like the contemporary frameworks \emph{Cinderella} and \emph{Cabri}, \jv\ moved on from the goals it had originally set. 
Other software solutions filled these gaps on the basis of more modern developments and now replace the interactive web elements that \jv\ once set out to create.
The \jv\ framework lives on as a stand-alone software that can be downloaded, installed, and run on a variety of platforms. 
Its user-friendly GUI and its XML file format are still available to the community. 

The discussion of \jv\ in this article is a mere case study and rather an example for challenges that can happen to any software framework. It shows that a visualization toolkit cannot safely focus on a unique selling point for an arbitrary amount of time. It must have either multiple use cases available---and be willing to drop one, should the need arise---or react timely to changes in its underlying architecture and its environment. These reactions include---if necessary---the translation into or combination with another language or the choice to move the whole framework into a different application area.

A gap can as easily arise in visualization software as in any situation where environmental factors are changing. We can try to learn from the presented examples, the historical developments, and the decisions that have been made in order to try and bridge gaps we are facing. Or we might come to the conclusion that---knowing the effort it takes to adjust a software system---it is not worth preserving it and it is more beneficial to move on to other approaches. Past developments can not take this decision from us, they can only guide us.

\section*{Acknowledgments}

The writing of this article has been supported by a Visiting Researcher Scholarship of the German National Academic Foundation and the Japanese RIKEN research institute. The author would like to thank the anonymous reviewers for their helpful comments and suggestions.



\bibliographystyle{plain}
\bibliography{ArXiv_literature.bib}

\end{document}